\documentclass[]{aa}
\usepackage{graphicx}
\usepackage{epsfig}
\usepackage{subfigure}
\usepackage{amssymb}
\usepackage{textgreek}
\usepackage{natbib}
\usepackage{subfigure}
\bibpunct{(}{)}{;}{a}{}{,}

\newcommand{\FIG}[1]{#1}
\def\mso{\,{\rm M}_\odot}
 \def\gcm{\,{\rm g}\,{\rm cm}^{-3}}
 \def\kms{\,{\rm km}\,{\rm s}^{-1}}
 \def\simle{\mathrel{\hbox{\rlap{\hbox{\lower4pt\hbox{$\sim$}}}\hbox{$<$}}}}
 \def\simgr{\mathrel{\hbox{\rlap{\hbox{\lower4pt\hbox{$\sim$}}}\hbox{$>$}}}}
 \def\vinf{\,\mathrm{v}_\infty}
 \def\mdot{\,\dot{M}}
 \def\msoy{\,\mso~{\rm yr}^{-1}}
 \def\muG{\,\textmu G}

\begin{document}

   \title{Can the magnetic field in the Orion arm inhibit the growth of instabilities in the bow shock of Betelgeuse?}

    \author{A. J. van Marle
          \inst{1}
          \and
          L. Decin
          \inst{1}
          \and
          Z. Meliani
          \inst{2,3}
          }

   \offprints{A. J. van Marle}

   \institute{Institute of Astronomy, KU Leuven, 
              Celestijnenlaan 200D, B-3001 Leuven, Belgium \\
              \email{AllardJan.vanMarle@ster.kuleuven.be} \\
             \email{Leen.Decin@ster.kuleuven.be}             
              \and          
             Observatoire de Paris, 
            5 place Jules Janssen 92195 Meudon, France \\
              \email{zakaria.meliani@obspm.fr}
           \and 
           APC, Universit{\'e} Paris Diderot, 
           10 rue Alice Domon et L{\'e}onie Duquet, 75205 Paris Cedex 13, France
} 

\date{Received <date> / Accepted <date>}

\abstract
{Many evolved stars travel through space at supersonic velocities, 
which leads to the formation of bow shocks ahead of the star where the stellar wind collides with the interstellar medium (ISM). 
\emph{Herschel} observations of the bow shock of \textalpha-Orionis show that the shock is almost free of instabilities, 
despite being, at least in theory, subject to both Kelvin-Helmholtz and Rayleigh-Taylor instabilities.}
{A possible explanation for the lack of instabilities lies in the presence of an interstellar magnetic field. 
We wish to investigate whether the magnetic field of the interstellar medium (ISM) in the Orion arm can inhibit the growth of instabilities in the bow shock of
\textalpha-Orionis.}
{We used the code MPI-AMRVAC to make magneto-hydrodynamic simulations of a circumstellar bow shock, using 
the wind parameters derived for \textalpha-Orionis and interstellar magnetic field strengths of $B\,=$\,1.4,\, 3.0,\, and 5.0\muG, 
which fall within the boundaries of the observed magnetic field strength in the Orion arm of the Milky Way.} 
{Our results show that even a relatively weak magnetic field in the interstellar medium can suppress the growth of Rayleigh-Taylor and Kelvin-Helmholtz instabilities,  
which occur along the contact discontinuity between the shocked wind and the shocked ISM.}
{The presence of even a weak magnetic field in the ISM effectively inhibits the growth of instabilities in the bow shock. 
This may explain the absence of such instabilities in the Herschel observations of \textalpha-Orionis.}

  \titlerunning{Magnetic smoothing of \textalpha-Ori's bow shock}
  \authorrunning{van Marle, Decin \& Meliani}


  \maketitle

%

\section{Introduction}
\begin{figure}
\FIG{
 \centering
\mbox{
\includegraphics[width=0.8\columnwidth]{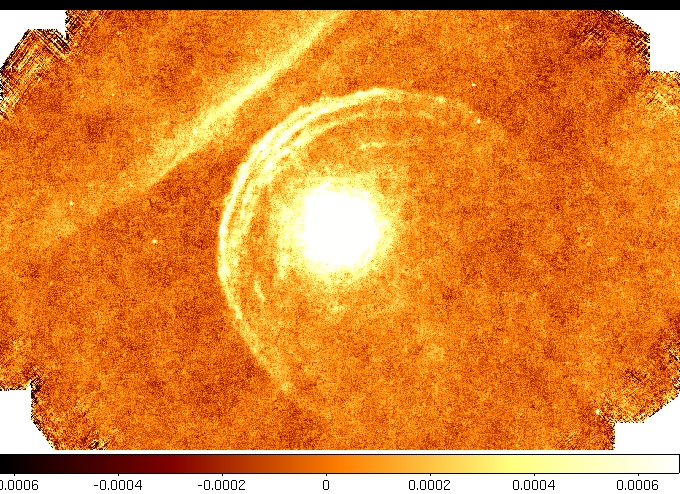}}
}
\caption{Herschel PACS image of the bow shock of \textalpha-Orionis at 70\,{\textmu m} in Jy/pixel for a field of view of 2275\,''$\times$1500\,''. The bow shock consists of multiple smooth arcs, 
that lack large-scale instabilities. 
}
 \label{fig:Herschel}\bf 
\end{figure}

Recent \emph{Herschel} observations \citep{Decinetal:2012} have given us a detailed view of the bow shock of \textalpha-Orionis, 
which is formed by the collision between the stellar wind and the interstellar medium (ISM). 
These observations show that the bow shock is smooth, without large instabilities 
(See Fig.~\ref{fig:Herschel} for the PACS observation of the bow shock at 70\,{\textmu m}. 
This traces the dust, rather than the gas, but \citet{vanMarleetaldust:2011} showed that small dust grains will follow the instabilities in the gas). 
Similar smooth shapes have been observed for the bow shocks of UU~Aur and X~Pav \citep{Coxetal:2012}. 
This runs is in contrast to both analytic predictions and numerical models of the wind-ISM collision. 
Based on analytic models, \citet{Dganietal:1996} predicted that a circumstellar bow shock is unstable when the space velocity of the star exceeds the velocity of 
the stellar winds, as is the case for \textalpha-Orionis. 
Numerical models by \citet{BrighentiDercole:1995} showed both Rayleigh-Taylor (RT) and Kelvin-Helmholtz (KH) instabilities 
for the general case. 
Subsequent attempts by \citet{vanMarleetaldust:2011} and \citet{Decinetal:2012} to model the bow shock of \textalpha-Orionis showed large instabilities 
that would be observable with \emph{Herschel}, if they exist.
The absence of instabilities has been explained by \citet{Mohamedetal:2012} and \citet{Mackeyetal:2012} as evidence that the bow shock is still in an early stage of formation and 
that the instabilities have not yet had time to develop. 

In this paper we explore an alternative theory: 
that the growth of instabilities is inhibited by an interstellar magnetic field. 
Interstellar magnetic fields are typically weak, but can stretch out over large distances \citep[$\simeq\,100$\,pc][]{RandKulkarni:1989,OhnoShibata:1993,Beck:2009,Shabalaetal:2010}.
Estimates for the magnetic field in the Orion~arm at a distance of 8\,000\,kpc from the Galactic centre (corresponding to the location of \textalpha-Orionis) 
range from $1.4\pm0.3$\muG\, \citep{Fricketal:2001} through 2-3\muG\, in the region near the sun \citep{HeerikhuisenPogorelov:2011} 
to 3.7-5.5\muG\, based on \emph{Voyager} measurements \citep{Opheretal:2009}. 
Similar values were obtained for the Galactic disk in general from \emph{WMAP} data \citep{JanssonFarrar:2012a,JanssonFarrar:2012b}. 

The influence of interstellar magnetic fields on the bow shocks of evolved stars was described analytically by \citet{Heiligman:1980} and \citet{SokerDgani:1997}. 
\citet{Keppensetal:1999} showed that a uniform magnetic field, aligned with the motion of the flow, can effectively inhibit the formation of KH instabilities.  
However, the conditions in the bow shock of \textalpha-Orionis are more complicated because the contact discontinuity is subject to RT instabilities as well.  

In this paper we present a 2.5-D model of the \textalpha-Orionis bow shock with and without an interstellar magnetic field. 
We based our input parameters for the stellar wind on the estimates by \citet{Uetaetal:2008} and introduce a magnetic field of 3.0\muG. 
This corresponds to the estimated field in the Orion~arm \citep{HeerikhuisenPogorelov:2011} as well as to the value found by \citet{Decinetal:2012} based on the separation of the arcs in the bow shock. 
It also conforms to typical field strengths found in the spiral arms of other galaxies \citep{Fletcheretal:2011,Vallee:2011}. 
We repeated the simulation for magnetic field strengths of 1.4\muG\, \citep[conform to][]{Fricketal:2001} and 5.0\muG\, \citep[conform to][]{Opheretal:2009} to investigate the quantitative effect of the field strength on the result. 

Animations of our results are shown in electronic format in Appendix~A.

\begin{table}[htp]
\centering
 \caption{Physical parameters of the wind of \textalpha-Orionis and the local ISM. }
\label{Table:input_AOri}
\begin{tabular}{lccl}
 \hline \hline
Mass loss rate                  & $\mdot$\,&=&\,$3.0\times10^{-6}$\,$\msoy$ \\
Wind velocity       & $\vinf$\,&=&\,15.0\,$\kms$\\
Velocity w.r.t.\ ISM    & $v_\star$\,&=&\,28.3\,$\kms$\\
ISM density            & $\rho_{\rm ISM}$\,&=&\,$10^{-23.5}\gcm$ \\
ISM temperature           & $T_{\rm ISM}$\,&=&\,10\,K \\
ISM magnetic field                  & $B$\,&=& 1.4, 3.0, 5.0\muG\, \\
\noalign{\smallskip}
\hline
\end{tabular}\\
\end{table}

\section{Numerical method}
For the simulations, we used the {\tt MPI-AMRVAC} magneto-hydrodynamics code \citep{vanderHolstetal:2008,Keppensetal:2012}. 
This is a fully conservative, finite-volume code that solves the conservation equations for mass, momentum and energy. 
For these simulations we chose to use ideal magneto-hydrodynamics and added optically thin radiative cooling, 
using the exact integration method \citep{Townsend:2009} and a cooling curve for interstellar gas at solar metallicity, 
generated with the code {\tt CLOUDY} \citep[][Wang Ye, private communication]{Ferlandetal:1998}. 

We simulated the interaction between the moving star and the ISM on a 2.5-D cylindrical grid of 160\,$\times$\,160\,grid cells, 
covering a physical domain of 2\,$\times$\,2\,pc in the r,z-plane. 
We allowed the adaptive mesh an additional four levels of refinement,  based on local variations in the absolute velocity of the gas, 
for an effective resolution of 2\,560\,$\times$\,2\,560 grid cells. 

The equations were solved using the total variation diminishing, Lax-Friedrich {\tt TVDLF} method combined with a koren flux limiter \citep{Kuzmin:2006}, 
except for the highest level of refinement, 
where we used the more diffusive minmod flux limiter \citep{Roe:1986} to prevent the growth of numerical instabilities at the polar axis.

We set the input conditions for our model according to the values in Table~\ref{Table:input_AOri}, with the stellar wind and ISM parameters based on \citet{Uetaetal:2008}.  
The stellar wind was simulated by filling a small sphere ($R$\,=\,0.1\,pc) with material according to the stellar wind parameters from Table~\ref{Table:input_AOri}, 
assuming a constant mass flux.  
The motion of the star through the ISM wass included by letting the ISM stream pass the star with a velocity $v_\star\,=\,28.3$\,km\,s$^{-1}$. 
We assumed the ISM to be cold, so the thermal pressure of the ISM is insignificant compared with the ram pressure created by the motion of the star. 

We run the simulation with magnetic field strengths of 0, 1.4, 3.0, and 5.0\,\muG, 
covering the parameter space of estimated magnetic field strengths in the Orion~arm of the Milky Way
\citep{Fricketal:2001,Opheretal:2009,HeerikhuisenPogorelov:2011}. 
The magnetic field was set parallel to the direction of motion and fixed at the upper z-boundary.
At the lower boundary we did not specify the magnetic field, because the tail of the bow shock, which crosses the boundary, will distort the field. 
We intend to investigate the effect of a magnetic field perpendicular to the direction of motion in 3D in the future.

\begin{figure*}
\FIG{
 \centering
\mbox{
\subfigure
{\includegraphics[width=0.33\textwidth]{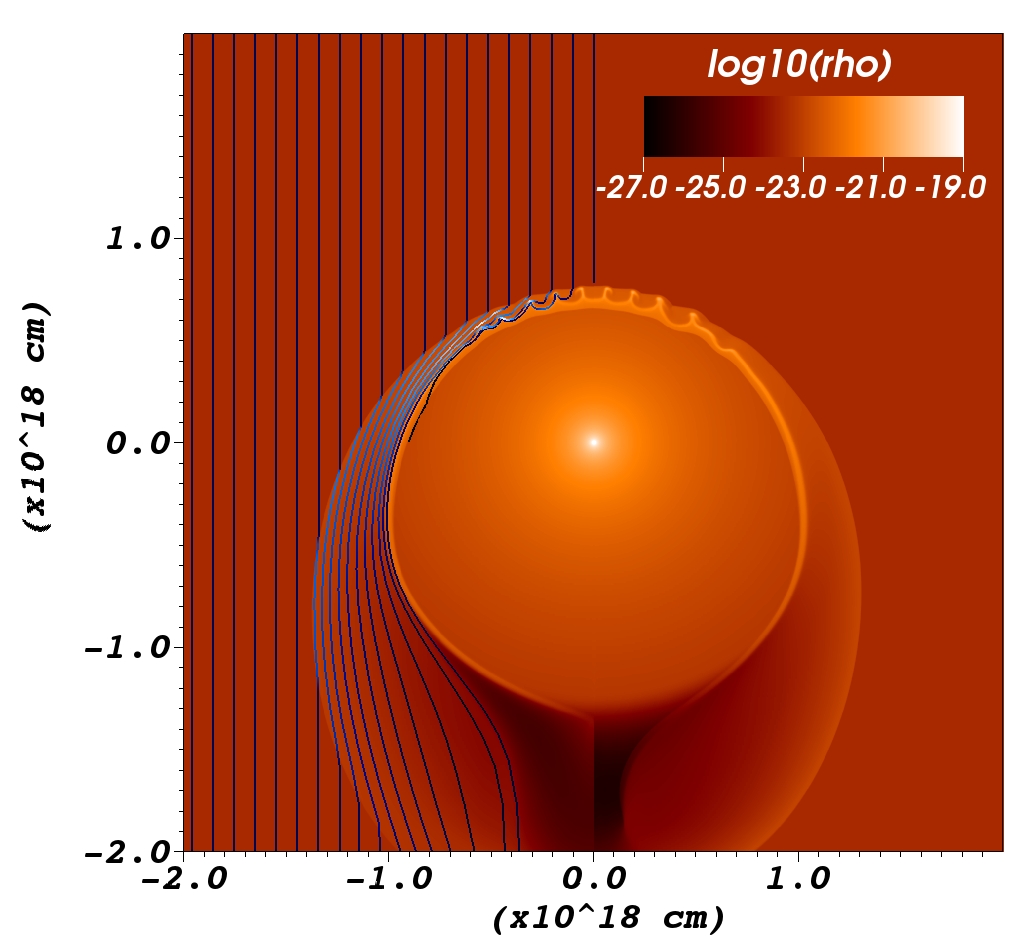}}
\subfigure
{\includegraphics[width=0.33\textwidth]{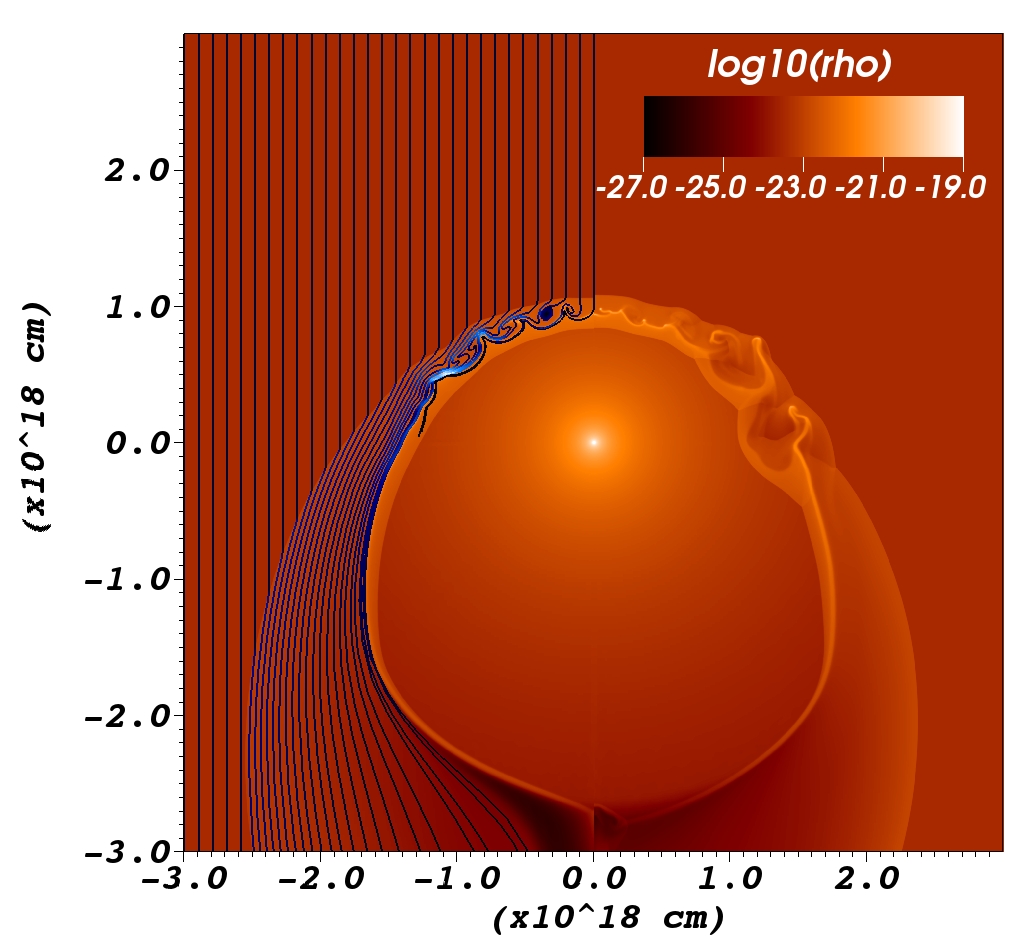}}
\subfigure
{\includegraphics[width=0.33\textwidth]{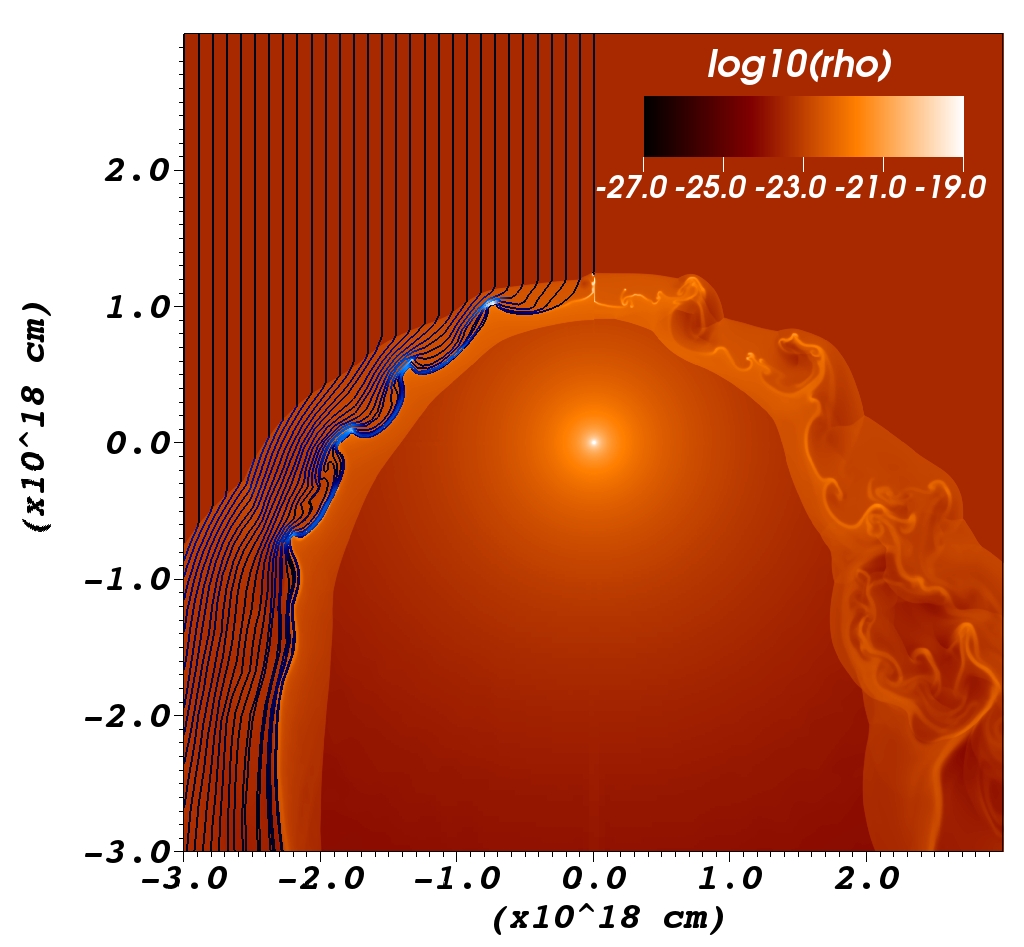}}}
}
\caption{Density of the circumstellar medium in $\gcm$ after 20\,000 (left), 50\,000 (centre) and 100\,000 (right) yr, for 
a simulation without an interstellar magnetic field (right side of each panel) and a 3\muG\, interstellar magnetic field (left side of each panel). 
The field lines for the latter model are also shown. 
In both simulations RT instabilities form initially at the contact discontinuity ahead of the star. 
For the non-magnetic model these instabilities grow as they travel downstream, distorting the shape of the bow shock. 
As they grow, they take on characteristic of KH instabilities. 
In the presence of the interstellar magnetic field the instabilities shrink as they move past the star,  and there is no sign of KH instabilities.  
} 
 \label{fig:Aori_B_1_3}
\end{figure*}

\section{Results}
\subsection{Non-magnetic vs. $B\,=\,3$\muG}
We used the intermediate magnetic field (3\muG) as a baseline for a comparison with the non-magnetic model (See Fig.~\ref{fig:Aori_B_1_3}). 
From an early phase (t\,$\geq$\,20\,000\,yrs, left panel of Fig.~\ref{fig:Aori_B_1_3}), the interaction between the wind and the ISM shows signs of instability.  
These instabilities start at the front of the bow shock (directly ahead of the star). 
For the non-magnetic model the instabilities grow as they travel downstream. 
In the presence of an interstellar magnetic field, their growth stops and is reversed. 

The difference between the two simulations becomes more pronounced over time. 
After 50\,000\, years (centre panel of Fig.~\ref{fig:Aori_B_1_3}), the non-magnetic model (right panel) shows large instabilities that combine the characteristics of both 
KH and RT instabilities. 
These instabilities grow in size as they move downstream, 
becoming large enough ($\simeq\,0.1$\,pc) to they distort the shape of both the reverse shock and the forward shock. 
In the magnetic model the instabilities stop growing as they move downstream, and the shape of the forward and reverse shocks is preserved. 
 
This pattern continues over time, as demonstrated in the right panel of Fig.~\ref{fig:Aori_B_1_3}, which shows the density after 100\,000\,years. 
At this time the bow shock has reached its equilibrium position determined by the ram pressure balance between the stellar wind and the ISM 
at a distance of approximately 0.3\,pc from the star \citep[e.g.][]{Uetaetal:2008,Decinetal:2012}. 
The non-magnetic bow shock is extremely unstable, showing a combination of KH and RT instabilities 
that deforms the smooth shape of both the wind termination shock and the forward shock. 
The bow shock of the magnetic model remains smooth, with only small RT instabilities 
and no sign of KH instabilities at all. 

\subsection{Dependence on magnetic field strength}
\begin{figure}
\FIG{
 \centering
\mbox{
\includegraphics[width=0.8\columnwidth]{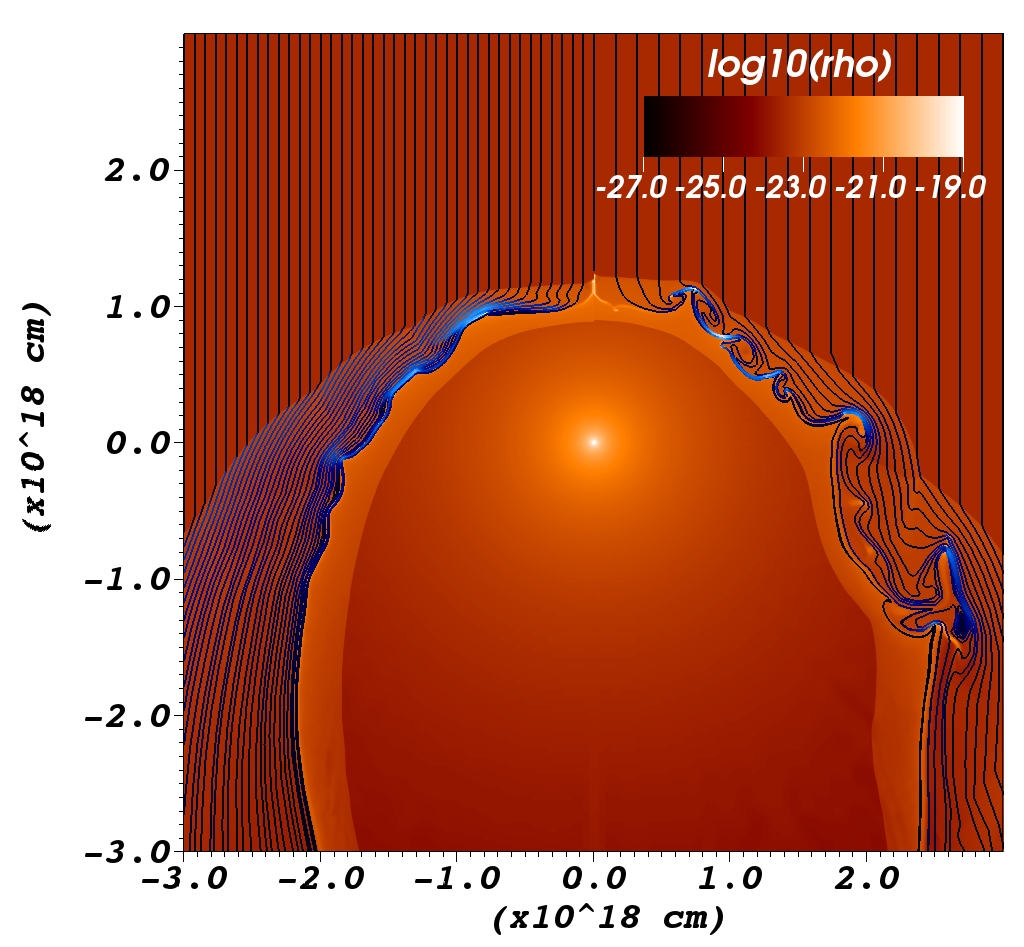}}
}
\caption{Similar to the right panel of Fig.~\ref{fig:Aori_B_1_3} for simulations with a 1.4\muG\, ISM field (right) and a 5.0\muG\, field (left) after 100\,000\,yr.  
Although the 1.4\muG\, field reduces the size and complexity of the instabilities, it is insufficient to stop them from growing. 
The 5.0\muG\, field, on the other hand, dampens the instabilities completely. 
}
 \label{fig:Aori_B_4}
\end{figure}

When we repeated the simulation for interstellar magnetic field strengths of $B\,=\,1.4$\muG\, and $B\,=\,5.0$\muG, we obtained the results shown 
in Fig.~\ref{fig:Aori_B_4}. 
The weak (1.4\,\muG) magnetic field, shown on the right, is clearly insufficient to stop the formation of either RT or KH instabilities, 
although the instabilities are more structured than for the model without magnetic field and their amplitude is reduced. 
Downstream, the field lines are twisted around by the KH instabilities to form the characteristic cat's eye shapes.
The strong (5.0\,\muG) field, shown on the left, almost completely suppresses all instabilities, to the point where they become all but invisible. 

\section{Nature of the instabilities}
For a bow shock without an interstellar magnetic field the contact discontinuity between the shocked wind is subject to two types of instabilities: 
owing to the velocity difference between the shocked wind and the shocked ISM, a shear force occurs that creates KH instabilities. 
At the front of the bow shock, the shocked wind, which was initially moving along the direction of motion of the star, 
has to make a turn to move downstream, causing it to experience a centrifugal force  \citep{BrighentiDercole:1995}. 
Because the less dense, shocked ISM decelerates the wind, the interface becomes RT unstable,  
with the centrifugal force performing the function that gravitational acceleration has in classical RT instabilities, 
but only along the curve.

Without a magnetic field, the RT instabilities at the front of the bow shock grow quickly and, by the time they have travelled past the star, are large enough 
to distort the shape of the bow shock, even though at this point the interface is no longer RT unstable. 
As they move further downstream the KH instability becomes increasingly prominent creating cat's eyes.

With the added influence of the interstellar magnetic field the situation becomes more complicated. 
Ahead of the star the magnetic field has to make a sharp curve as it enters the bow shock, creating a magnetic tension vector that works against 
the motion of the RT instabilities. 
The small wavelength perturbations are suppressed even by the weakest (1.4\muG) magnetic field. 
The stronger magnetic fields also reduce the amplitude of the instabilities at longer wavelengths. 
The RT instability cannot be completely suppressed, because 
it will effectively be moving along the field lines in the shocked ISM. 
Still, the growth rate of the instabilities is reduced and they do not expand perpendicular to their original direction to form the characteristic mushroom shape. 

Downstream of the star, the magnetic field runs parallel with the interface. 
As demonstrated by \citet{Keppensetal:1999}, this effectively prevents the formation of KH instabilities.
\citet{Stonegardiner:2007} showed that RT instabilities can still occur under these conditions. 
However, in this region the interface is no longer RT unstable. 
Magnetic pressure on the interface can create an interchange instability. 
However, the magnetic pressure ($3.5\,\times\,10^{-13}$\,dyn\,cm$^{-2}$ for a 3\muG\, field) is much lower than 
the ram pressure of the wind or the stellar motion into the ISM ($2.5\,\times\,10^{-11}$\,dyn\,cm$^{-2}$ when the gas enters the bow shock).  
The Alfv{\'e}n speed in the unshocked ISM is 4.6$\kms$ for the 3\muG\, field, slower than either the wind or the shocked ISM. 
Therefore the growth time of a magnetically induced instability is likely to be longer than the dynamical time scale of the bow shock. 
This follows the analytical prediction and numerical simulation by \citet{Junetal:1995} for the contribution of the magnetic field toward counteracting RT instabilities. 
Moreover, in our model the magnetic field is tangential, which increases the stability and suppresses the linear and non-linear growth of the RT instability. 
For a uniform tangential magnetic field, the growth rate is given by 
\begin{equation}
n^2=k g\frac{\rho_1-\rho_2}{\rho_1+\rho_2}-B^2\,k^2\frac{\cos^2\theta}{2\,\pi\left(\rho_1+\rho_2\right)}\,,
\end{equation}
with $n$ the growth rate, $k$ the wave number, $g$ the acceleration at the contact discontinuity, 
$\rho_1$ and $\rho_2$ the shocked ISM and wind densities, respectively, $B$ the magnetic field strength, and $\theta$ the angle between the $\vec{B}$ and $\vec{k}$. 

\section{Conclusions}
Our simulations show that even a relatively weak magnetic field, 
parallel to the direction of motion, effectively inhibits the growth of instabilities in the bow shock. 
Even the intermediate field (3.0\muG), which falls well within the boundaries of the estimated field strength for the Orion~arm,  
can explain why the \emph{Herschel} observations of \textalpha-Orionis  (see Fig.~\ref{fig:Herschel}) show a smooth bow shock \citep{Decinetal:2012}  
despite the instability of the hydrodynamical interaction. 

Simulations reported in \citet{Decinetal:2012} showed that a warm ISM ($\sim\,8000\,$K) can reduce the size of the instabilities, even without the presence of a magnetic field. 
We can assume that a combination of an interstellar magnetic field with a warm ISM would reduce the instabilities 
even further than either can do on their own. 

Although the influence of the magnetic field on the behaviour of the local instabilities is profound, 
it does not influence the general size or shape of the bow shock.  
The bow shock, supported by the ram pressure of the stellar wind, bends the magnetic field lines away from the star,  
and the field is not strong enough to overcome this force, though it is strong enough to suppress the instabilities.

It is impossible to simulate the effect of a magnetic field perpendicular to the direction of motion on a two dimensional grid 
because this field configuration lacks a symmetry axis. 
However, we can tentatively predict the effect based on our results: 
on a large scale, the motion of the star will compress the magnetic field, increasing its strength. 
This would constrain the radius of the bow shock. 
The field would inhibit the formation of KH instabilities in the region ahead of the star. 
Moreover, the field lines would curve around the bow shock \citep{Dganietal:1996}, so the inhibiting effect can be expected to continue in the downstream region. 
Farther downstream, the contact discontinuity would still be KH unstable. 
However, this region falls outside the smooth bow-shock region observed with \emph{Herschel} (see Fig.~\ref{fig:Herschel}). 
The behaviour of RT instabilities under these conditions would be more complicated and has to be investigated in 3-D \citep{Stonegardiner:2007}.

In the future we plan to use 3D simulations to investigate the effect of the perpendicular magnetic field as well 
as the influence of magnetic fields of different strength and magnetic fields with a turbulent component. 

\begin{acknowledgements} 
A.J.v.M.\ acknowledges support from FWO, grant G.0277.08, K.U.Leuven GOA/2008/04 and GOA/2009/09. 
We thank Wang Ye at the Department of Physics \& Astronomy, University of Kentucky, for providing us with the radiative-cooling curve. 
PACS was developed by a consortium of institutes led by the MPE (Germany) and including UVIE (Austria); KUL, CSL, IMEC (Belgium); CEA, OAMP (France); 
MPIA (Germany); IFSI, OAP/AOT, OAA/CAISMI, LENS, SISSA (Italy); IAC (Spain). The development has been supported by the funding agencies BMVIT (Austria), ESA-PRODEX (Belgium), 
CEA/CNES (France), DLR (Germany), ASI (Italy), and CICT/MCT (Spain).
\end{acknowledgements}

\bibliographystyle{aa}
\bibliography{vanmarle_biblio}

\begin{thebibliography}{30}
\expandafter\ifx\csname natexlab\endcsname\relax\def\natexlab#1{#1}\fi

\bibitem[{{Beck}(2009)}]{Beck:2009}
{Beck}, R. 2009, Astrophysics and Space Sciences Transactions, 5, 43

\bibitem[{{Brighenti} \& {D'Ercole}(1995)}]{BrighentiDercole:1995}
{Brighenti}, F. \& {D'Ercole}, A. 1995, \mnras, 277, 53

\bibitem[{{Cox} {et~al.}(2012){Cox}, {Kerschbaum}, {van Marle}, {Decin},
  {Ladjal}, {Mayer}, {Groenewegen}, {van Eck}, {Royer}, {Ottensamer}, {Ueta},
  {Jorissen}, {Mecina}, {Meliani}, {Luntzer}, {Blommaert}, {Posch},
  {Vandenbussche}, \& {Waelkens}}]{Coxetal:2012}
{Cox}, N.~L.~J., {Kerschbaum}, F., {van Marle}, A.-J., {et~al.} 2012, \aap,
  537, A35

\bibitem[{{Decin} {et~al.}(2012){Decin}, {Cox}, {Royer}, {Van Marle},
  {Vandenbussche}, {Ladjal}, {Kerschbaum}, {Ottensamer}, {Barlow}, {Blommaert},
  {Gomez}, {Groenewegen}, {Lim}, {Swinyard}, {Waelkens}, \&
  {Tielens}}]{Decinetal:2012}
{Decin}, L., {Cox}, N.~L.~J., {Royer}, P., {et~al.} 2012, \aap, 548, A113

\bibitem[{{Dgani} {et~al.}(1996){Dgani}, {van Buren}, \&
  {Noriega-Crespo}}]{Dganietal:1996}
{Dgani}, R., {van Buren}, D., \& {Noriega-Crespo}, A. 1996, \apj, 461, 927

\bibitem[{{Ferland} {et~al.}(1998){Ferland}, {Korista}, {Verner}, {Ferguson},
  {Kingdon}, \& {Verner}}]{Ferlandetal:1998}
{Ferland}, G.~J., {Korista}, K.~T., {Verner}, D.~A., {et~al.} 1998, \pasp, 110,
  761

\bibitem[{{Fletcher} {et~al.}(2011){Fletcher}, {Beck}, {Shukurov},
  {Berkhuijsen}, \& {Horellou}}]{Fletcheretal:2011}
{Fletcher}, A., {Beck}, R., {Shukurov}, A., {Berkhuijsen}, E.~M., \&
  {Horellou}, C. 2011, \mnras, 412, 2396

\bibitem[{{Frick} {et~al.}(2001){Frick}, {Stepanov}, {Shukurov}, \&
  {Sokoloff}}]{Fricketal:2001}
{Frick}, P., {Stepanov}, R., {Shukurov}, A., \& {Sokoloff}, D. 2001, \mnras,
  325, 649

\bibitem[{{Heerikhuisen} \& {Pogorelov}(2011)}]{HeerikhuisenPogorelov:2011}
{Heerikhuisen}, J. \& {Pogorelov}, N.~V. 2011, \apj, 738, 29

\bibitem[{{Heiligman}(1980)}]{Heiligman:1980}
{Heiligman}, G.~M. 1980, \mnras, 191, 761

\bibitem[{{Jansson} \& {Farrar}(2012{\natexlab{a}})}]{JanssonFarrar:2012a}
{Jansson}, R. \& {Farrar}, G.~R. 2012{\natexlab{a}}, \apj, 757, 14

\bibitem[{{Jansson} \& {Farrar}(2012{\natexlab{b}})}]{JanssonFarrar:2012b}
{Jansson}, R. \& {Farrar}, G.~R. 2012{\natexlab{b}}, \apjl, 761, L11

\bibitem[{{Jun} {et~al.}(1995){Jun}, {Norman}, \& {Stone}}]{Junetal:1995}
{Jun}, B.-I., {Norman}, M.~L., \& {Stone}, J.~M. 1995, \apj, 453, 332

\bibitem[{{Keppens} {et~al.}(2012){Keppens}, {Meliani}, {van Marle}, {Delmont},
  {Vlasis}, \& {van der Holst}}]{Keppensetal:2012}
{Keppens}, R., {Meliani}, Z., {van Marle}, A.~J., {et~al.} 2012, Journal of
  Computational Physics, 231, 718

\bibitem[{{Keppens} {et~al.}(1999){Keppens}, {T{\'o}th}, {Westermann}, \&
  {Goedbloed}}]{Keppensetal:1999}
{Keppens}, R., {T{\'o}th}, G., {Westermann}, R.~H.~J., \& {Goedbloed}, J.~P.
  1999, Journal of Plasma Physics, 61, 1

\bibitem[{{Kuzmin}(2006)}]{Kuzmin:2006}
{Kuzmin}, D. 2006, Journal of Computational Physics, 219, 513

\bibitem[{{Mackey} {et~al.}(2012){Mackey}, {Mohamed}, {Neilson}, {Langer}, \&
  {Meyer}}]{Mackeyetal:2012}
{Mackey}, J., {Mohamed}, S., {Neilson}, H.~R., {Langer}, N., \& {Meyer},
  D.~M.-A. 2012, \apjl, 751, L10

\bibitem[{{Mohamed} {et~al.}(2012){Mohamed}, {Mackey}, \&
  {Langer}}]{Mohamedetal:2012}
{Mohamed}, S., {Mackey}, J., \& {Langer}, N. 2012, \aap, 541, A1

\bibitem[{{Ohno} \& {Shibata}(1993)}]{OhnoShibata:1993}
{Ohno}, H. \& {Shibata}, S. 1993, \mnras, 262, 953

\bibitem[{{Opher} {et~al.}(2009){Opher}, {Bibi}, {Toth}, {Richardson},
  {Izmodenov}, \& {Gombosi}}]{Opheretal:2009}
{Opher}, M., {Bibi}, F.~A., {Toth}, G., {et~al.} 2009, \nat, 462, 1036

\bibitem[{{Rand} \& {Kulkarni}(1989)}]{RandKulkarni:1989}
{Rand}, R.~J. \& {Kulkarni}, S.~R. 1989, \apj, 343, 760

\bibitem[{{Roe}(1986)}]{Roe:1986}
{Roe}, P.~L. 1986, Annual Review of Fluid Mechanics, 18, 337

\bibitem[{{Shabala} {et~al.}(2010){Shabala}, {Mead}, \&
  {Alexander}}]{Shabalaetal:2010}
{Shabala}, S., {Mead}, J.~M.~G., \& {Alexander}, P. 2010, ArXiv e-prints

\bibitem[{{Soker} \& {Dgani}(1997)}]{SokerDgani:1997}
{Soker}, N. \& {Dgani}, R. 1997, \apj, 484, 277

\bibitem[{{Stone} \& {Gardiner}(2007)}]{Stonegardiner:2007}
{Stone}, J.~M. \& {Gardiner}, T. 2007, Physics of Fluids, 19, 094104

\bibitem[{{Townsend}(2009)}]{Townsend:2009}
{Townsend}, R.~H.~D. 2009, \apjs, 181, 391

\bibitem[{{Ueta} {et~al.}(2008){Ueta}, {Izumiura}, {Yamamura}, {Nakada},
  {Matsuura}, {Ita}, {Tanab{\'e}}, {Fukushi}, {Matsunaga}, \&
  {Mito}}]{Uetaetal:2008}
{Ueta}, T., {Izumiura}, H., {Yamamura}, I., {et~al.} 2008, \pasj, 60, 407

\bibitem[{{Vall{\'e}e}(2011)}]{Vallee:2011}
{Vall{\'e}e}, J.~P. 2011, \nar, 55, 91

\bibitem[{{van der Holst} {et~al.}(2008){van der Holst}, {Keppens}, \&
  {Meliani}}]{vanderHolstetal:2008}
{van der Holst}, B., {Keppens}, R., \& {Meliani}, Z. 2008, Computer Physics
  Communications, 179, 617

\bibitem[{{van Marle} {et~al.}(2011){van Marle}, {Meliani}, {Keppens}, \&
  {Decin}}]{vanMarleetaldust:2011}
{van Marle}, A.~J., {Meliani}, Z., {Keppens}, R., \& {Decin}, L. 2011, \apjl,
  734, L26

\end{thebibliography}

\IfFileExists{vanmarle_biblio.bbl}{}
 {\typeout{}
  \typeout{******************************************}
  \typeout{** Please run "bibtex \jobname" to obtain}
  \typeout{** the bibliography and then re-run LaTeX}
  \typeout{** twice to fix the references!}
  \typeout{******************************************}
  \typeout{}
 }

\listofobjects

\Online

\begin{appendix}  
\section{Animations of the bow shock evolution}
\begin{figure}
\FIG{
 \centering
\mbox{
\includegraphics[width=0.8\columnwidth]{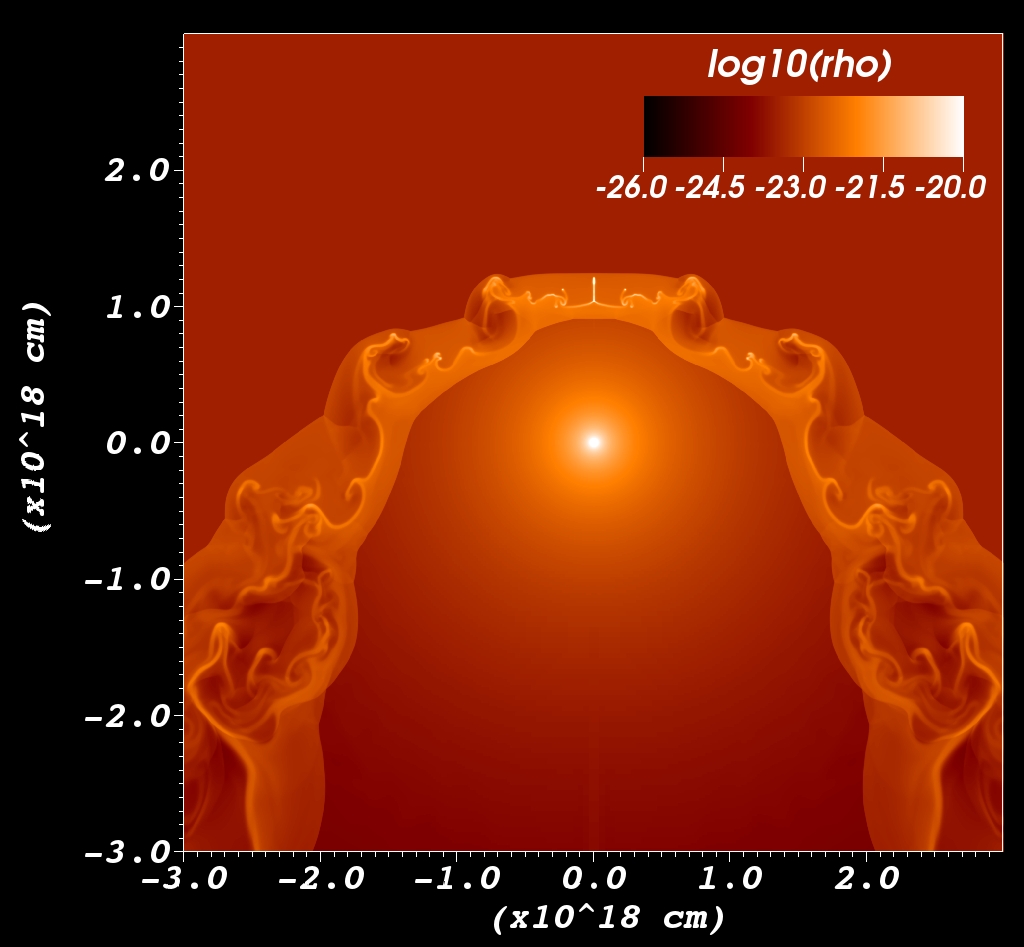}}
}
\caption{
This movie shows the density of the bow shock for a non-magnetic ISM.}
\end{figure}

\begin{figure}
\FIG{
 \centering
\mbox{
\includegraphics[width=0.8\columnwidth]{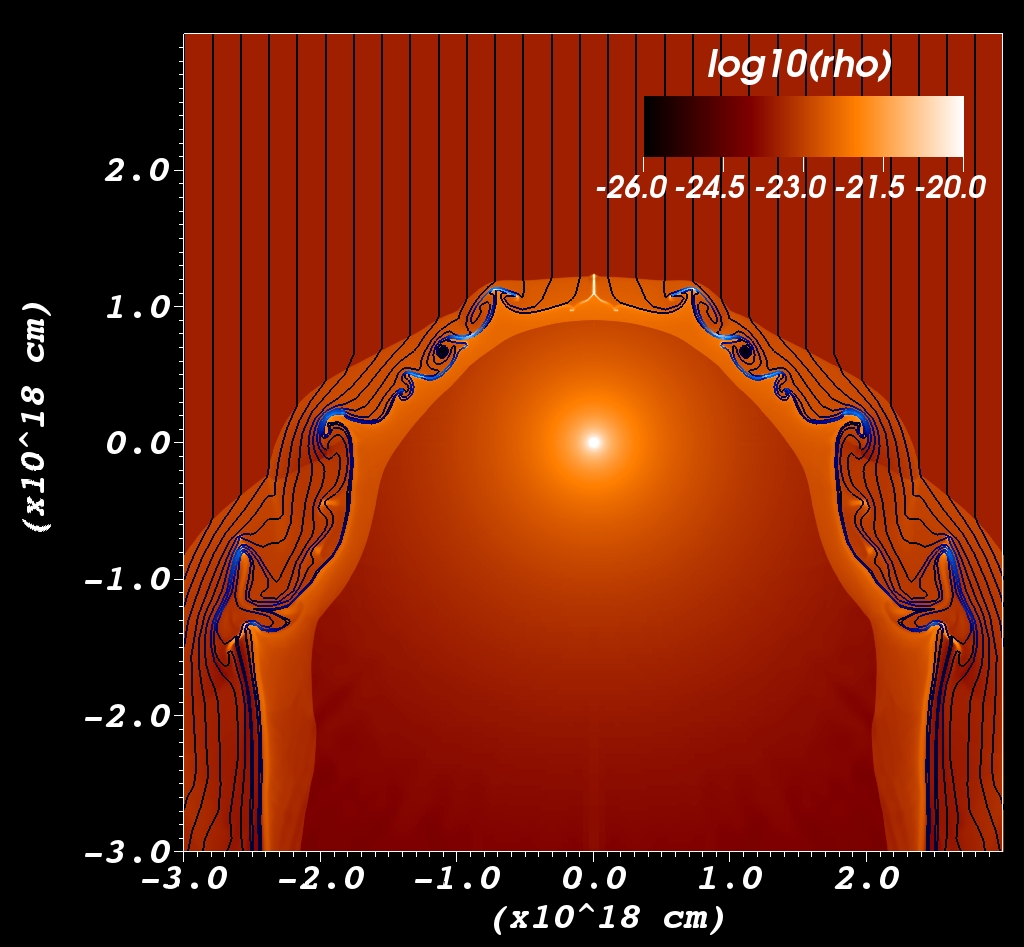}}
}
\caption{
This movie shows the density, as well as the magnetic field lines, of the bow shock for an ISM with a 1.4\muG\, magnetic field.}
\end{figure}

\begin{figure}
\FIG{
 \centering
\mbox{
\includegraphics[width=0.8\columnwidth]{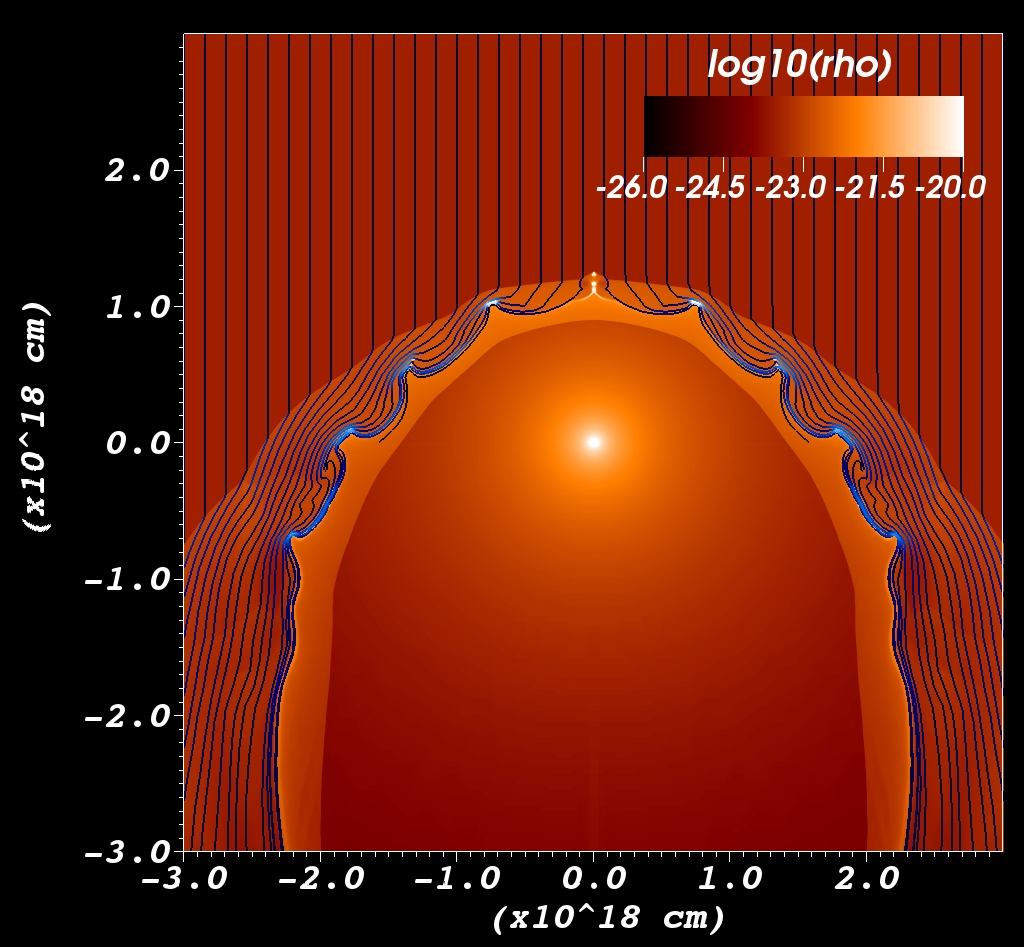}}
}
\caption{
This movie shows the density of the bow shock, as well as the magnetic field lines, for an ISM with a 3.0\muG\, magnetic field.}
\end{figure}

\begin{figure}
\FIG{
 \centering
\mbox{
\includegraphics[width=0.8\columnwidth]{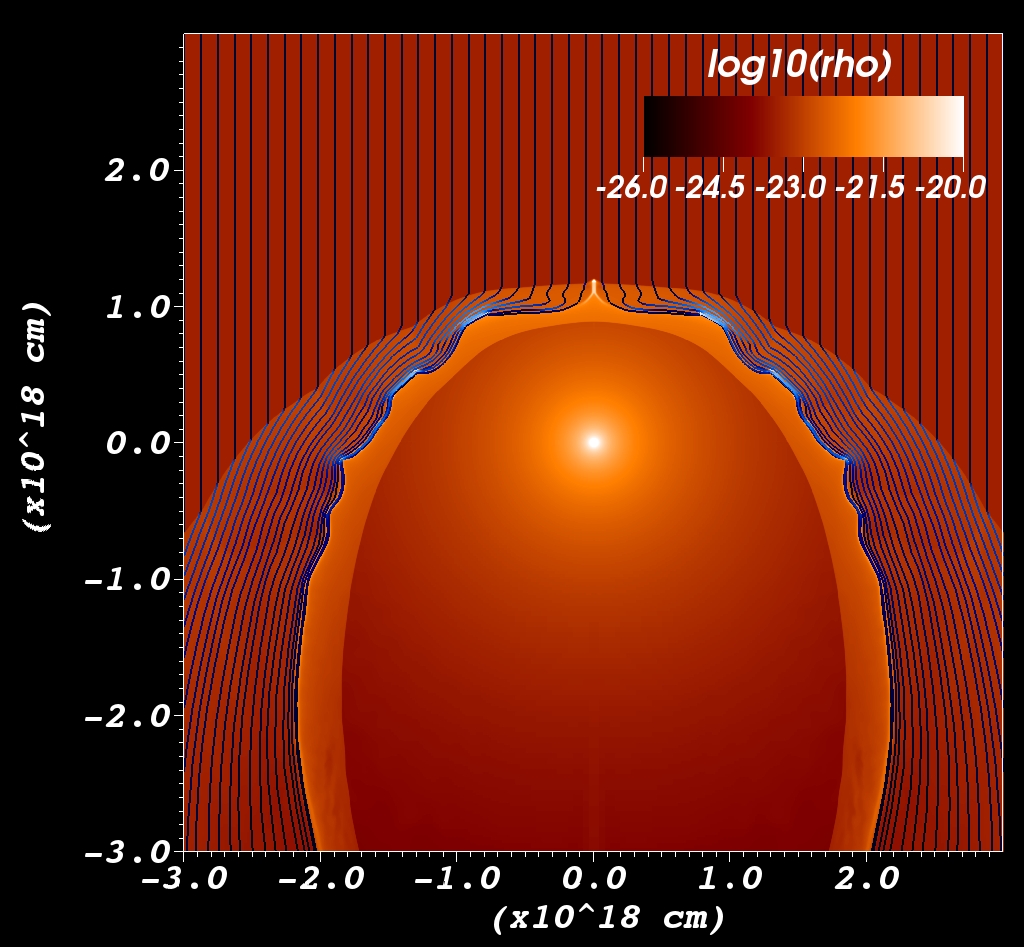}}
}
\caption{
This movie shows the density of the bow shock, as well as the magnetic field lines, for an ISM with a 5.0\muG\, magnetic field.}
\end{figure}

\end{appendix}
\end{document}